\documentclass[11pt]{spieman}  
\usepackage{amsmath,amsfonts,amssymb}
\usepackage{graphicx}
\usepackage{setspace}
\usepackage{tocloft}
\usepackage[square,sort,comma,numbers]{natbib}

\title{The Far-Infrared Enhanced Survey Spectrometer (FIRESS) for PRIMA: Science Drivers \footnote{\copyright 2025. All rights reserved. }}

\author[a]{Klaus M. Pontoppidan}
\author[b]{Alberto Bolatto}
\author[c]{J D Smith}
\author[a]{C. M. (Matt) Bradford}
\author[d]{Cara Battersby}
\author[e]{Alexandra Pope}
\author[a]{Tiffany Kataria}
\author[f]{Jason Glenn}
\author[a]{Margaret Meixner}

\author[g]{Lee Armus}
\author[h]{Jochem Baselmans}
\author[r]{Edwin A. Bergin}
\author[i]{Denis Burgurella}
\author[i]{Laure Ciesla}
\author[q]{L. Ilsedore Cleeves}
\author[j]{Anna Di Giorgio}
\author[j]{Carlotta Gruppioni}
\author[k]{Thomas Henning}
\author[a]{Brandon Hensley}
\author[l]{Willem Jellema}
\author[k]{Oliver Krause}
\author[m]{Elisabeth Mills}
\author[n]{Arielle Moullet}
\author[o]{Marc Sauvage}
\author[p]{Rachel Somerville}
\author[f]{Johannes Staguhn}
\author[a]{Steve Unwin}

\affil[a]{Jet Propulsion Laboratory, California Institute of Technology, 4800 Oak Grove, Pasadena, CA, USA, 91109}
\affil[b]{Department of Astronomy, University of Maryland, College Park, MD}
\affil[c]{Dept. of Physics and Astronomy, RO 3000A, The University of Toledo, Toledo, OH 43606, USA }
\affil[d]{Department of Physics, University of Connecticut, 196A Auditorium Rd, Unit 3046 Storrs, CT 06269-3046, USA}
\affil[e]{Department of Astronomy, University of Massachusetts, Amherst, MA 01003, USA }
\affil[f]{Goddard Space Flight Center, 8800 Greenbelt Road, Greenbelt, MD, USA 20771}
\affil[g]{Infrared Processing and Analysis Center, Caltech, MC 314-6, 1200 E California Blvd, Pasadena, CA 91125 USA }
\affil[h]{TU Delft, Fac. EEMCS, Mekelweg 4, 2628 CD Delft, Netherlands \& SRON Leiden, Niels Bohrweg 4, 2333 CA Leiden, Netherlands}
\affil[i]{Aix Marseille Univ, CNRS, CNES,  Laboratoire d'Astrophysique de Marseille,  Technopole Chateau-Gombert, 38, rue Frederic Joliot-Curie, 13388 Marseille cedex 13, France}
\affil[j]{Istituto Nazionale di Astrofisica: Osservatorio di Astrofisica e Scienza dello Spazio di Bologna, Via Gobetti 93/3, 40129 Bologna, Italy}
\affil[k]{Max Planck Institute for Astronomy (MPIA) Königstuhl 17, 69117 Heidelberg, Germany}
\affil[l]{SRON Groningen, Landleven 12, 9747 AD Groningen, Netherlands}
\affil[m]{Dept. of Physics and Astronomy, University of Kansas, Malott Hall, 1251 Wescoe Hall Dr., Lawrence, KS 66045}
\affil[n]{National Radio Astronomy Observatory (NRAO), 520 Edgemont Road, Charlottesville, VA 22903 USA}
\affil[o]{Universit\'e Paris-Saclay, Univ. Paris Cit\'e, CEA, CNRS, AIM, 91191, Gif-sur-Yvette, France}
\affil[p]{Center for Computational Astrophysics, Flatiron Institute, 162 5th Avenue, New York, NY 10010, USA}
\affil[q]{University of Virginia, Department of Chemistry, Charlottesville, VA 22904, USA}
\affil[r]{Department of Astronomy, University of Michigan, 1085 South University Avenue, Ann Arbor, MI 48109, USA}

\cftpagenumbersoff{figure}
\cftpagenumbersoff{table} 
\begin{document} 
\maketitle

\begin{abstract}
We present the science drivers for the Far-Infrared Enhanced Survey Spectrometer (FIRESS), one of two science instrument on the PRobe Infrared Mission for Astrophysics (PRIMA). FIRESS is designed to meet science objectives in the areas of the origins of planetary atmospheres, the co-evolution of galaxies and supermassive black holes, and the buildup of heavy elements in the Universe. In addition to these drivers, FIRESS is envisioned as a versatile far-infrared spectrometer, capable of addressing science questions in most areas of astrophysics and planetary astronomy as part of a dominant General Observer (GO) program with 2/3 of the current science cases using FIRESS. We summarize how the instrument design choices and parameters enable the main science drivers as well as a broad and vibrant GO program. 
\end{abstract}


{\noindent \footnotesize\textbf{*}Klaus Pontoppidan,  \linkable{klaus.m.pontoppidan@jpl.nasa.gov} }


\section{Introduction}
\label{sect:intro}
The PRobe Infrared Mission for Astrophysics (PRIMA) takes advantage of the maturation of key technologies to understand the origins of galaxies, stars, and planets, by providing sensitive access to the far-infrared waveband (24-235\,$\mu$m), resulting in improvements in survey speed of 3-5 orders of magnitude compared to the previous state-of-the art \citep[][\S\ref{section:speed}]{Glenn25}. It is at these wavelengths that we may measure key tracers of dust, ice, and molecular gas, and where the universe emits half its light across the past 10 billion years of its history \citep[The Cosmic Infrared Background, CIB;][]{Fixsen98,Hill18}. PRIMA is a 1.8\,m actively cooled telescope feeding two science instruments. The PRIMA Imager (PRIMAger) offers hyperspectral imaging at 24-80\,$\mu$m and polarimetric broad-band imaging in four bands spanning 80-262\,$\mu$m \citep{Ciesla25}. 

The second PRIMA science instruments, the Far-Infrared Enhanced Survey Spectrometer \citep[FIRESS,][]{Bradford25} offers a versatile set of spectroscopic observing modes, including spectral mapping, and deep point-source spectroscopy. The PRIMA focal planes employ low-noise state-of-the-art microwave kinetic inductance detectors (MKIDs) enabling broad wavelength coverage and fast survey speeds. This combination offers sensitive access to the far-infrared wavelength range, limited by the astrophysical background rather than by telescope self-emission or detector noise. FIRESS is equipped with a total of 672 spectral channels, half of which (336) can observe a point source simultaneously, enabling a full-band, low-resolution ($R\sim 100$) grating spectrum to be obtained in just two settings. The addition of a pre-dispersed Fourier Transform Module produces a high-resolution ($R\sim 4400\times 112\,\mu \mathrm{m}/\lambda$) spectrum that provides simultaneous access to the same spectral band as the low-resolution mode.

FIRESS succeeds previous space-borne spectrometers on the Infrared Space Observatory \citep[ISO, ][]{Kessler96}, the Herschel Space Observatory \citep{Pilbratt10}, the Stratospheric Observatory for Infrared Astronomy \citep[SOFIA, ][]{Becklin07}, and the long-wavelength edges of the Spitzer Space Telescope \citep{Werner04} and the James Webb Space Telescope \citep[JWST, ][]{Gardner06}. While these historical missions made transformational advances relative to previous capabilities, they were sensitivity-limited by telescope background emission and/or detector noise. PRIMA's cold optics, low-noise Kinetic Inductance Detectors, and large-format arrays enables observations at the astrophysical background limit allowing us to address a wide range of astrophysical questions beyond the reach of any previous far-infrared facility. It is this leap in spatial and spectral multiplexing, and therefore spectral mapping speed of FIRESS, that provides access to science objectives not previously possible. For instance, the Long-Wavelength Spectrometer \citep[LWS, ][]{Clegg96, Swinyard96} on ISO, launched in 1998, was limited by available technology at the time to only 10 doped germanium photoconductors to provide access to a 47-196\,$\mu$m spectral grasp. This led to spectral observing modes that required time-consuming scans in the spectral direction, thus limiting sensitive observations to individual lines. Later, the Photoconductor Array Camera and Spectrometer \citep[PACS, ][]{Poglitsch10} onboard Herschel was equipped with photoconductor arrays, offering more individual pixels enabling the deployment of a 5x5 spaxel integral field unit, but was ultimately still limited to narrow spectral range scans for the observation of faint lines offering 16 spectral pixels, with its sensitivity limited by both detector noise (NEP$\sim 0.9-2.1\times 10^{-18}\,\rm W\,Hz^{-1/2}$) and the warm telescope ($\sim 70\,$K). This mode was used for the deep spectroscopy that for the first time detected hydrogen deuteride (HD) in three protoplanetary disks to measure their masses and mapped large areas within nearby galaxies in key gas cooling lines of [CII], [OI], and [NII] to reveal the heating/cooling balance of the ISM, but at great cost in observing time, and only yielding a single line in a narrow bandpass. Because FIRESS is designed to offer fast, broad-band observing speeds at the astrophysical limit, there would be less advantage in very high spectral resolving powers. FIRESS is therefore highly complementary to instruments offering single-line $R>>10,000$ spectroscopy, such as SOFIA-GREAT \citep{Heyminck12}, or the upcoming Planetary Origins and Evolution Multispectral Monochrometer (POEMM) balloon mission.

This paper describes the key science drivers for FIRESS, how they flow to instrument requirements, and how FIRESS enables a wide range of General Observer science. It follows on the FIRESS technical description in \citep{Bradford25}.

\section{Key science drivers for FIRESS}

PRIMA-FIRESS is a versatile, multi-mode instrument capable of addressing a wide range of questions across much of astrophysics. While the instrumental design of FIRESS is driven by a limited set of science objectives, the derived technical capabilities are designed to enable and support many science cases. The FIRESS design is defined based on three basic science objectives. 

\begin{itemize}
\item Determine the volatile element budgets for oxygen and carbon in the primary gas reservoirs for the formation of planets, measured using broad-band, high resolution  spectroscopy of a large number of water lines in protoplanetary disks, as well as the ground state line ($J=1-0$) of hydrogen deuteride (HD) at 112\,$\mu$m to measure the total disk masses (Section \ref{sec:planet}). 

\item Determine the relation between the growth rates of stars and supermassive black holes, by combining large hyperspectral imaging surveys obtained with PRIMAger with a FIRESS-measured relation between classical line tracers and the far-infrared spectral energy distribution. Further, determine the role of galactic winds in the evolution of SMBHs and their host galaxies by measuring the wind mass-loss rates using high-resolution spectroscopy of far-infrared hydroxyl (OH) lines (Section \ref{sec:coevol}) .
\item Understand the formation and buildup of heavy elements and interstellar dust in the universe by measuring the relation between the [OIII] and [NIII] fine structure lines and Polycyclic Aromatic Hydrocarbon (PAH) emission in galaxies at cosmic noon and beyond (Section \ref{sec:buildup}). 
\end{itemize}

Since 75\% of PRIMA's 5-year lifetime is dedicated to General Observer (GO) programs, FIRESS is capable of addressing a wide range of science questions across most areas of astrophysics and planetary astronomy. The PRIMA General Observer (GO) book \citep{Moullet23} collected 76 science cases, of which 2/3 requested FIRESS either alone or in combination with PRIMAger, the other PRIMA science instrument. FIRESS accesses the full bandpass in all its observing modes (mapping modes, low- and high resolving power), and is therefore well-suited to address new science questions, requiring far-infrared spectroscopy, that may emerge in the 2030s. We discuss this potential throughout this paper and specifically in Section \ref{sec:discover}.

\begin{figure}[ht]
\includegraphics[width=\textwidth]{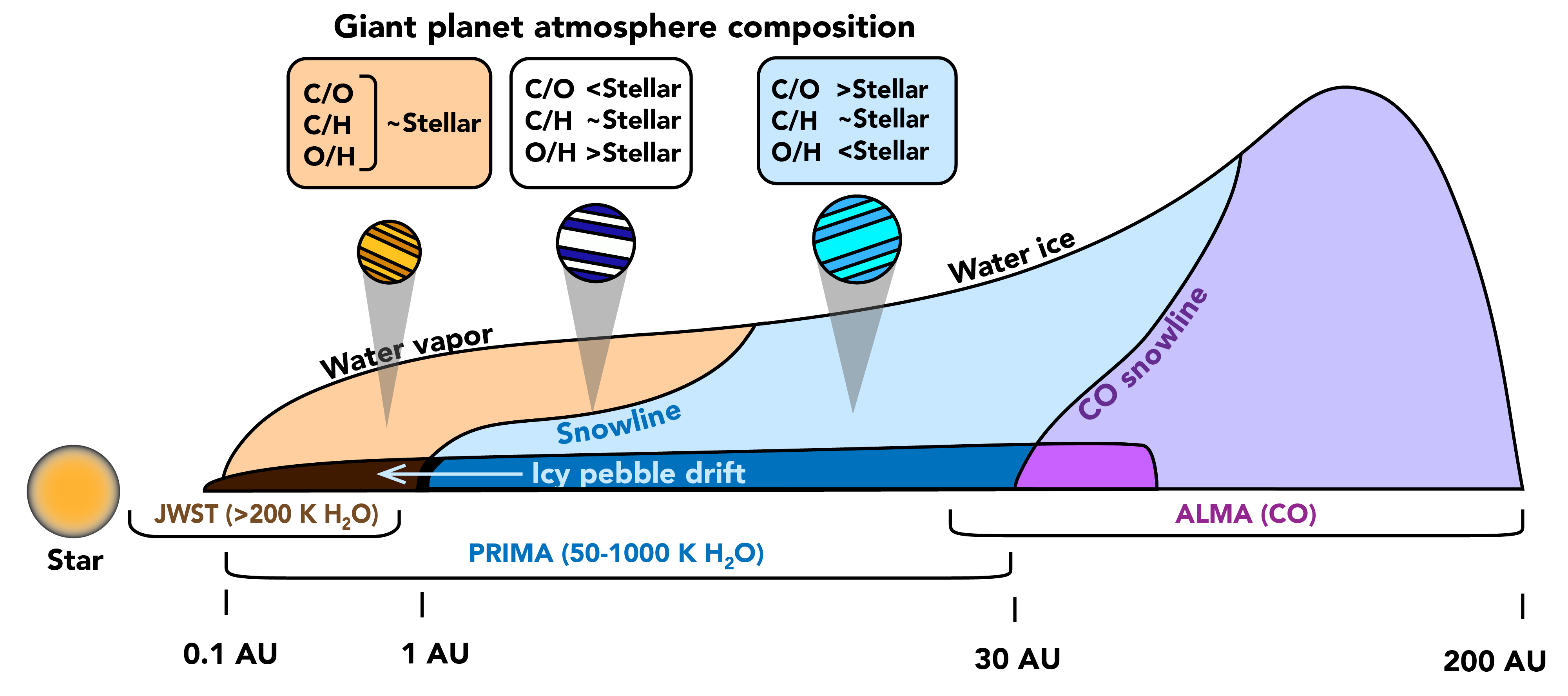}
\caption{The bulk composition of planets reflect that of their natal protoplanetary disks at the disk radii where planets accrete their mass. For rocky planets, the bulk composition will be that of the solids of the disk, whereas giant planet atmospheres derive their composition from the disk gas. The total abundance of the disk ([C/H] and [O/H]) as well as the carbon-to-oxygen ratio (C/O) as measured by PRIMA-FIRESS are key tracers of how disk chemical evolution shapes planetary atmospheres.}
\label{fig:disk_cartoon}
\end{figure}

\subsection{The origins of planetary atmospheres}
\label{sec:planet}
While planet cores develop from the solid reservoir of a protoplanetary disk, giant planet atmospheres develop from the gas reservoir \citep{Mordasini16}. Consequently, carbon and oxygen abundances ([C/H], [O/H], C/O) in transiting exoplanet atmospheres have been hypothesized to reflect their abundances in the disk during planet formation \citep[][Figure \ref{fig:disk_cartoon}]{Oberg11,Madhusudhan12}. We currently rely on theoretical models \citep{Eistrup18} to link these abundances \citep[][]{Bean18,Constantinou23} because accurate total disk gas masses are not available. 

Water is believed to play a pivotal role in this process since it dominates the solid mass outside the water “snowline” (the location in the disk where water condenses into solid grains). Water ice forms the basis of pebble-sized (mm to cm) icy grains, whose radial drift (icy pebble drift) concentrates solid mass in protoplanetary disks \citep{Ros13} and catalyzes the rapid formation of planetesimals and planetary cores \citep{Lambrechts19,Pinilla20}. ALMA observations indicate that this pebble drift may be ubiquitous in outer disks ($>$30 AU) \citep{Pinilla12, Appelgren25}, but its role in the inner disk ($<$30 AU) where most planets form is largely unconstrained, and only now being studied with JWST \citep{Banzatti23,Romero-Mirza24}. Icy pebbles drifting inwards through the snowline at $\sim$150\,K may deplete the outer disk in water ice, evaporate and enrich the inner disk with water vapor, thereby changing the bulk disk C/O ratio as a function of radius and evolutionary stage \citep{Colmenares24,Kanwar24,Long25}, as well as the relative mass distribution between inner and outer disk volatile reservoirs\citep{Houge25}. FIRESS directly measures the radial distribution of water vapor to constrain the importance of icy pebble drift and its relation to planetary architecture and atmospheric composition.

Another unknown is the total disk gas mass as a function of disk age, which sets the time scale for planet formation, and measures the denominator in the volatile abundances [O/H] and [C/H]. PRIMA obtains the total H$_2$ disk mass from the 112\,$\mu$m HD ground-state line and uses it as a key input parameter to global thermochemical disk models (e.g., Dust And LInes (DALI), \citep{Bruderer12}) to determine the “H” in [C/H] and [O/H]. HD is $\sim 10^6$ times more emissive than H$_2$, the bulk gas reservoir in disks, and has a well-established HD/H$_2$ ratio limited by the elemental D/H abundance \citep{Bergin19}, which is known to $\sim$30\% precision \citep{Linsky98}. The “O” in [O/H] is from PRIMA measurements of water, while the “C” in [C/H] is known from existing ALMA observations of CO. We compare disk abundances that are oxygen-enriched inside their snowline from icy pebble drift with those that are carbon-depleted from freeze-out outside the snowline, in order to determine how disk composition influences planet formation.

\begin{figure}[ht]
\centering
\includegraphics[width=10cm]{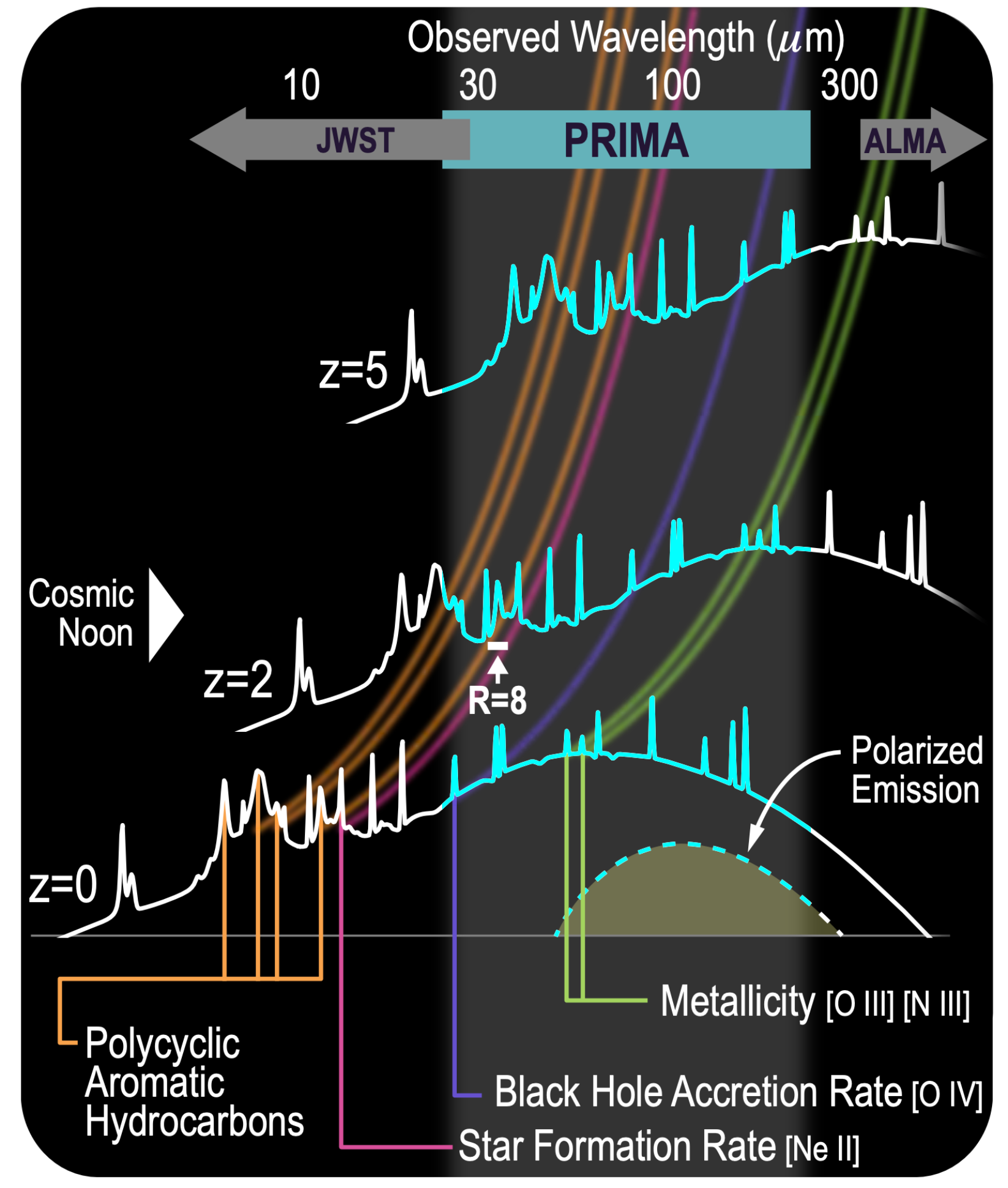}
\caption{FIRESS is capable of measuring the abundances of PAHs at redshifts $z=2-5$ and unobscured tracers of star formation and black hole accretion rates at $z>1$. }
\label{fig:galaxy_tracers}
\end{figure}

\subsection{The co-evolution of galaxies and black holes}
\label{sec:coevol}
Every massive galaxy hosts a supermassive black hole (SMBH) at its center. The SMBH mass is tightly correlated with the galaxy bulge mass and with its total stellar mass \citep{Reines2015}, suggesting that galaxies and their SMBHs grow together with evolutionary processes shaping both the galaxies and their central black holes. The most rapid growth of galaxies and their SMBHs happens during a period of heavy obscuration by dust and gas, around the peak of cosmic comoving star formation rate density at $z\sim2$ known as ``Cosmic Noon'' \citep{Zavala2021}. Obscuration by a combination of the Active Galactic Nucleus (AGN) torus, circumnuclear starbursts, and large-scale galaxy material is a major barrier to characterize the co-evolution of galaxies and their central black holes during rapid growth \citep{Hickox2018}. Rest-frame hard X-ray emission can be an effective probe of AGN \citep[e.g.,][]{Koss2011}, although at very high column densities ($N_H>1.5\times10^{24}$\,cm$^{-2}$) Compton interactions absorb even 10 keV photons \citep{Comastri2004}, and the fastest growing black holes accreting at super-Eddington rates are most likely intrinsically faint in X-rays
\citep{Begelman1979,Pacucci2024}. 

However, to accurately constrain the co-evolution of galaxies and their black holes it is necessary to measure both the star formation and black hole accretion rates in the same galaxies for large samples. The far-infrared spectral energy distribution (SED) provides straightforward access to both parameters, with the luminosity of the obscured AGN concentrated in a hot component contributing at rest-frame mid-infrared ($10-25$\,$\mu$m) wavelengths, and the star formation-related luminosity in a cool component taking over at longer wavelengths \citep{Kirkpatrick2013,Shimizu2017}. Thus, A powerful approach to simultaneously obtaining an accurate census of highly obscured AGN along with the host galaxies' star formation activity during ``Cosmic Noon'' is using relatively unobscured light received at far-infrared wavelengths and the fast survey speed of PRIMA to gather information for very large samples ($>$10,000 galaxies out to $z\gtrsim 2$). FIRESS plays an important role by accurately verifying the SED-based rates using higher-resolution spectroscopy for a subset of galaxies of rest-frame mid- to far-infrared fine-structure lines tracing both star formation ([Ne II]) and SMBH accretion rates ([O IV]). There are only a few complete 30-250\,$\mu$m far-infrared spectra of nearby galaxies ever obtained, and none at high redshift. FIRESS can readily fill this gap for significant samples of galaxies across a wide range of redshifts (See Figure \ref{fig:galaxy_tracers}). 

Winds containing large masses of gas, ejected from active black holes into the circumgalactic medium, are a key feedback mechanism linking SMBH and galaxy growth. Models invoke these winds to quench star formation in massive galaxies because they provide a mechanism to both remove star-forming gas and to prevent new gas from accreting \citep{Mitchell22}. However, model predictions are fundamentally constrained by the lack of measured wind properties such as their velocities and mass outflow rates \citep{Nelson19}. Observations of infrequent or weak winds would change the paradigm. The cool (molecular and neutral atomic, $T\lesssim 10^4$\,K) wind component dominates the outflowing mass, providing the most relevant outflow rates. Existing measurements of this component are limited to small samples ($<20$) of local galaxies \citep[e.g.,][]{Veilleux20}. PRIMA measures cool galactic winds in active galaxies out to $z\sim 2.0$.

\begin{figure}[ht]
\centering
\includegraphics[width=10cm,angle=270]{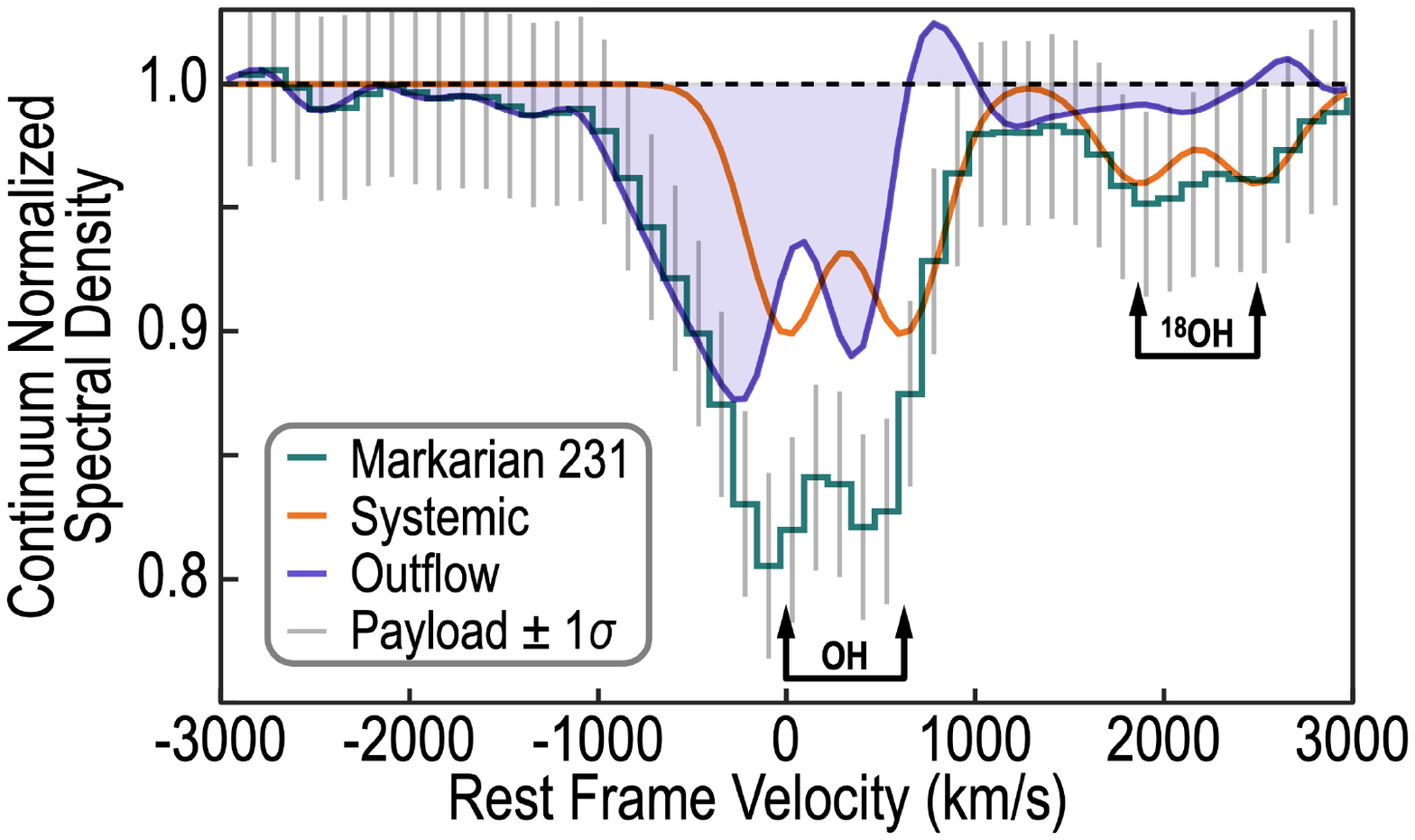}
\caption{Simulation of a FIRESS high-resolution observation of OH absorption from a galactic wind. In this case, the resolving power is tuned to $R=900$ to optimize the sensitivity to the line profile.}
\label{fig:oh}
\end{figure}

\subsection{The Buildup of Heavy Elements and Dust}
\label{sec:buildup}
Since the first massive stars began exploding as supernovae after cosmic dawn, the Universe has produced a growing complement of heavy elements. Heavy elements play a fundamental role in the formation of galaxies, stars, and planetary systems. When heavy elements are injected into the interstellar medium (ISM) by asymptotic giant branch (AGB) stars (both C and O-rich stars) and supernovae \citep{Matsuura09}, a large fraction is converted to dust—a near-universal step in their complex journey to forming planetary systems, including our own solar nebula \citep{Nittler03}. The remarkably efficient growth and evolution of dust from heavy elements over cosmic time remains a mystery \citep{Boyer12}. PRIMA provides data in the form of key far-IR tracers of dust and metals at two pivotal times in the Universe: cosmic noon ($z=2$), when most metals formed and the cycle of dust production and destruction also peaks \citep{Popping17,Peroux20}; and the present day ($z=0$). PRIMA’s access to the far-IR spectral range directly addresses two unsolved, yet crucial, questions about the co-evolution of dust and metals. First, how has the relationship between gas-phase heavy-element contents (metallicity) and the mass fraction of small-grain dust traced with PAHs in galaxies changed since cosmic noon (z=2)? PRIMA can measure the PAH-metallicity relationship at cosmic noon because PAH bands have redshifted into the far-IR. 

Spitzer observations have shown that the smallest grains of carbon dust, PAHs, contribute up to 25\% of the IR luminosity of galaxies \citep{Smith07} and are ubiquitous from today until cosmic noon \citep{Langeroodi23}, and possibly beyond \citep{Markov23, Spilker23}. PAH feature strengths relative to the nearby dust continuum are used to measure the small-grain abundance ($q_{\rm PAH}$), which is suppressed in low-metallicity systems in the local universe \citep{Engelbracht05,Engelbracht08,Whitcomb23}. The main explanations include destruction of PAHs in harsh radiation environ- ments, and inhibited grain regrowth inside clouds with fewer free metals. The importance of this effect at the critical epoch when most heavy elements were formed remains unclear.

\section{Design philosophy}

The PRIMA PI science objectives are accomplished by three basic science observing modes with FIRESS; low-resolution mapping spectroscopy, low-resolution point-source spectroscopy, and high-resolution point-source spectroscopy. The incoming beam is split by dichroics into four wavelength bands, masked by long slits, and dispersed by four gratings with resolving powers of $R\sim 85-150$ \citep{Bradford25}. This design allows for rapid observation for point sources of the full spectral band from 24-235\,$\mu$m. Spectral mapping is accomplished by scanning the slits across the sky using flexible two-dimensional patterns using a beam-steering mirror (BSM) while continuously reading out four detector arrays. For the low-resolution point-source mode, the source is chopped between different slit positions using the BSM. To provide a higher-resolution mode, it is possible to insert a Fourier Transform Module (FTM) into the dispersed beam to provide separate interferograms for every grating spectral pixel. In this case, the source is nodded within the same slit with the BSM fixed. 

\begin{figure}[ht]
\centering
\includegraphics[width=17cm]{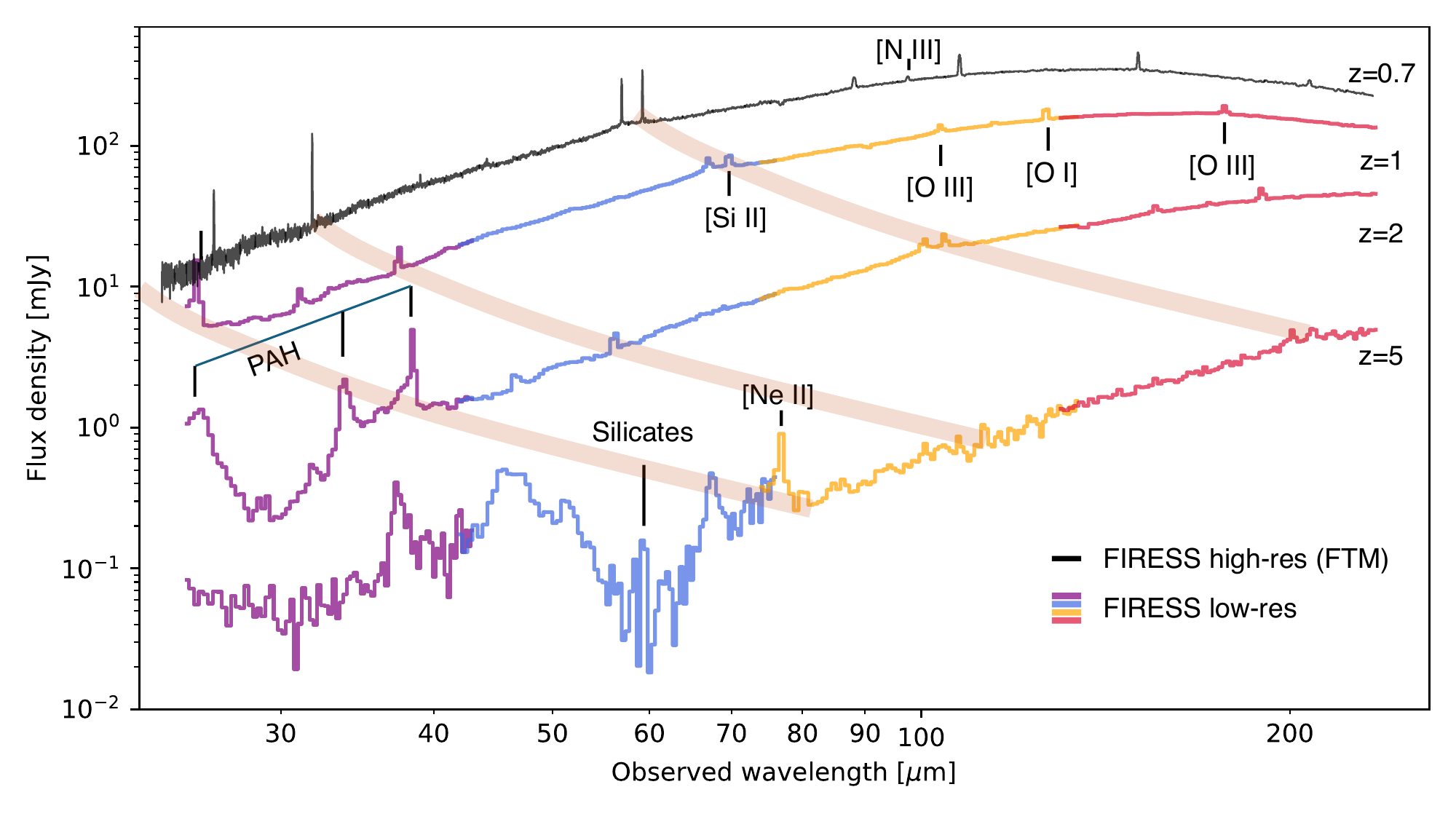}
\caption{Simulated FIRESS low-resolution observations of luminous ($10^{13.0}\,L_{\odot}$) starburst galaxies at high redshift, calculated by scaling the spectrum of M82, as observed with ISO-SWS and LWS \citep{Colbert99,Sturm00} (one of only a few complete far-infrared spectra of external galaxies). Representative noise realizations are added for exposure times per setting of 4, 2, 4, and 12 hours per setting for the z=0.7, 1, 2, and 5 galaxy, respectively; FIRESS requires two spectral settings to cover the entire wavelength range. The simulated high-resolution spectrum using the FIRESS FTM mode uses a (selectable) resolving power of R=600 at 112\,$\mu$m.}
\label{fig:galaxy spectra}
\end{figure}

\subsection{Spectral range}
FIRESS is designed to provide full, near-simultaneous spectral coverage in all its modes, requiring only two spectral settings to cover the full PRIMA band between 24 and 235\,$\mu$m. This is needed to access the same galactic star-formation and supermassive black hole (SMBH) tracers over a wide range of redshifts, and to observe a large number of water lines in protoplanetary disks. 

The short end of the wavelength coverage at 24\,$\mu$m enables measurements of the 12.8\,$\mu$m [Ne II] line, which traces star formation in galaxies at $z\gtrsim 1$, and ensures overlap with JWST-MIRI (which extends to $\sim$28\,$\mu$m). The JWST overlap is important for cross-calibration when combining JWST and PRIMA observations, especially for young stars, protoplanetary disks, and other sources with potential for significant variability. FIRESS also provides easy access to the ground-state H$_2$ S(0) line at 28.2\,$\mu$m, which is generally inaccessible to JWST, but is an important tracer of warm bulk molecular gas and the energy balance of molecular clouds \citep{Kaufman06}. The full range (24--235\,$\mu$m) allows for detections of strong 11.3\,$\mu$m PAH features at all relevant redshifts beyond $z \gtrsim 1.1$ and the 84\,$\mu$m OH doublet out to $z\sim 1.8$. 

\begin{figure}[ht]
\includegraphics[width=\textwidth]{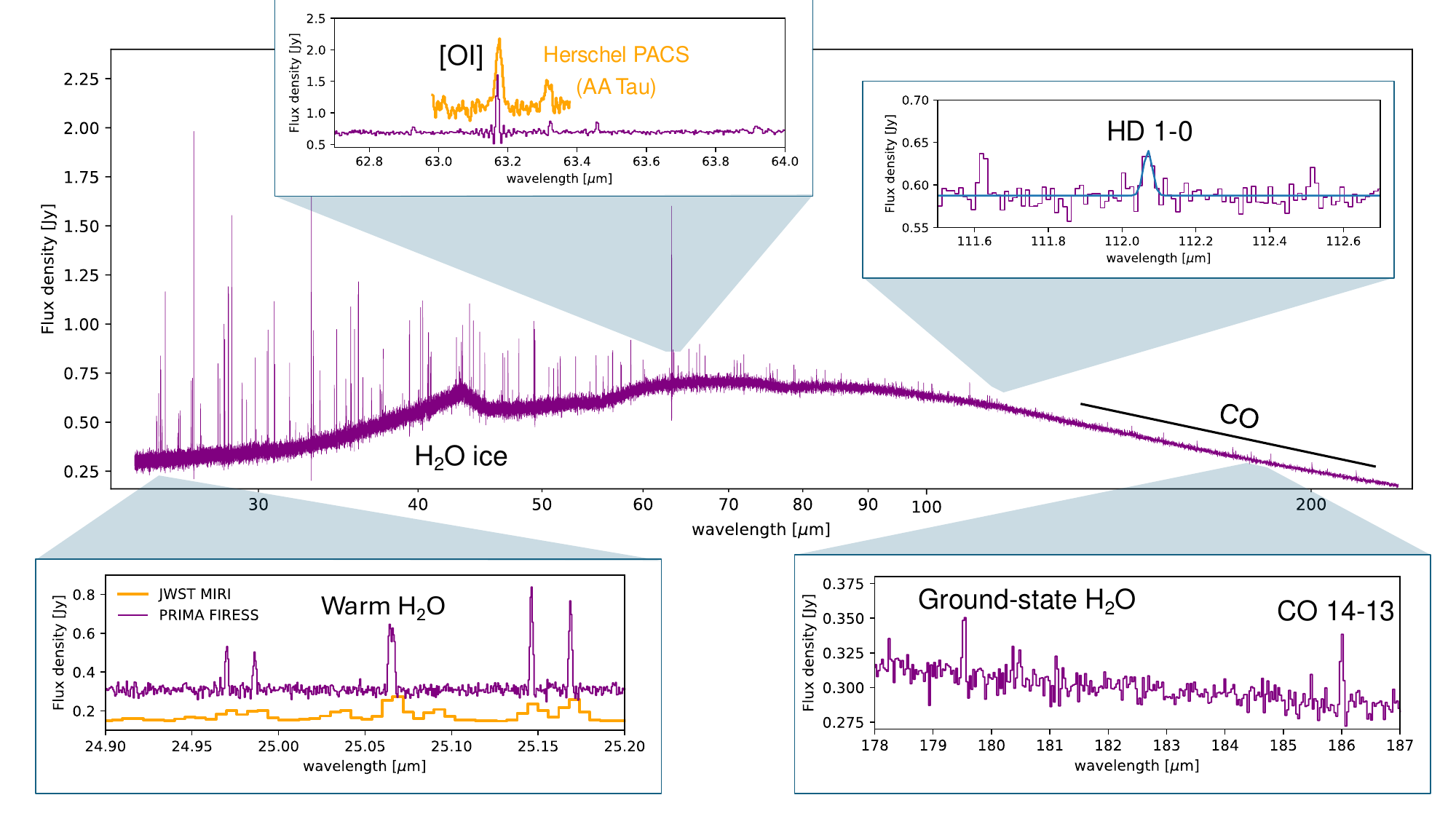}
\caption{Simulation of a full-band FIRESS spectrum of a generic protoplanetary disk with properties typical of those expected for young protoplanetary disk around a relatively low-mass star ($\sim 0.5\,M_{\odot}$). The noise is set by the required FIRESS sensitivity and assumes a total integration time of 10 hours per spectral setting and a (selectable) resolving power of $R=4400$ at 112\,$\mu$m. In this case, the HD line has an intensity of $4\times 10^{-19}\,\rm W\,m^{-2}$, and is detected with a signal-to-noise ratio of 10. The simulation is compared with JWST-MIRI \citep{Pontoppidan24} and Herschel PACS \citep{Blevins16} observations of similar disks. }
\label{fig:keystone}
\end{figure}

For protoplanetary disks, the range encompasses water lines of a wide range of upper level energies, including the ortho ground-state water line at 179.53\,$\mu$m (H$_2$O $2_{1 2}\rightarrow {1_{0 1}}$, $E_{\rm upper}=114.4\,$K), the 234.8\,$\mu$m $2_{12}\rightarrow 1_{01}$ HDO line ($E_{\rm upper}=83.6\,$K), lines from warm CO (J=12-11 at 217\,$\mu$m is the lowest-lying $^{12}$CO line accessible to FIRESS), and a wide range of atomic and ionic fine-structure lines from volatile elements (including 63 and 145 \,$\mu$m [OI], 158\,$\mu$m [C II], and 205\,$\mu$m [N II]. 

Finally, the FIRESS range provides access to important solid-state features, including the 43\,$\mu$m band of water ice (see Figure \ref{fig:keystone}), which has not been accessible since the mid-1990s with ISO \citep{Malfait99}, and the 69\,$\mu$m forsterite band, which is very sensitive to dust temperature and composition, but could only be effectively surveyed in protoplanetary disks around luminous intermediate-mass stars with Herschel \citep{Sturm13}. In the context of ices, the far-infrared is unique in that it traces ice in emission, and therefore has the potential to measure bulk ice mass in disks, young stars, and molecular clouds \citep{Kamp18}. 

\subsection{Spatial resolution and field of view}

\begin{figure}[ht]
\centering
\includegraphics[width=17cm]{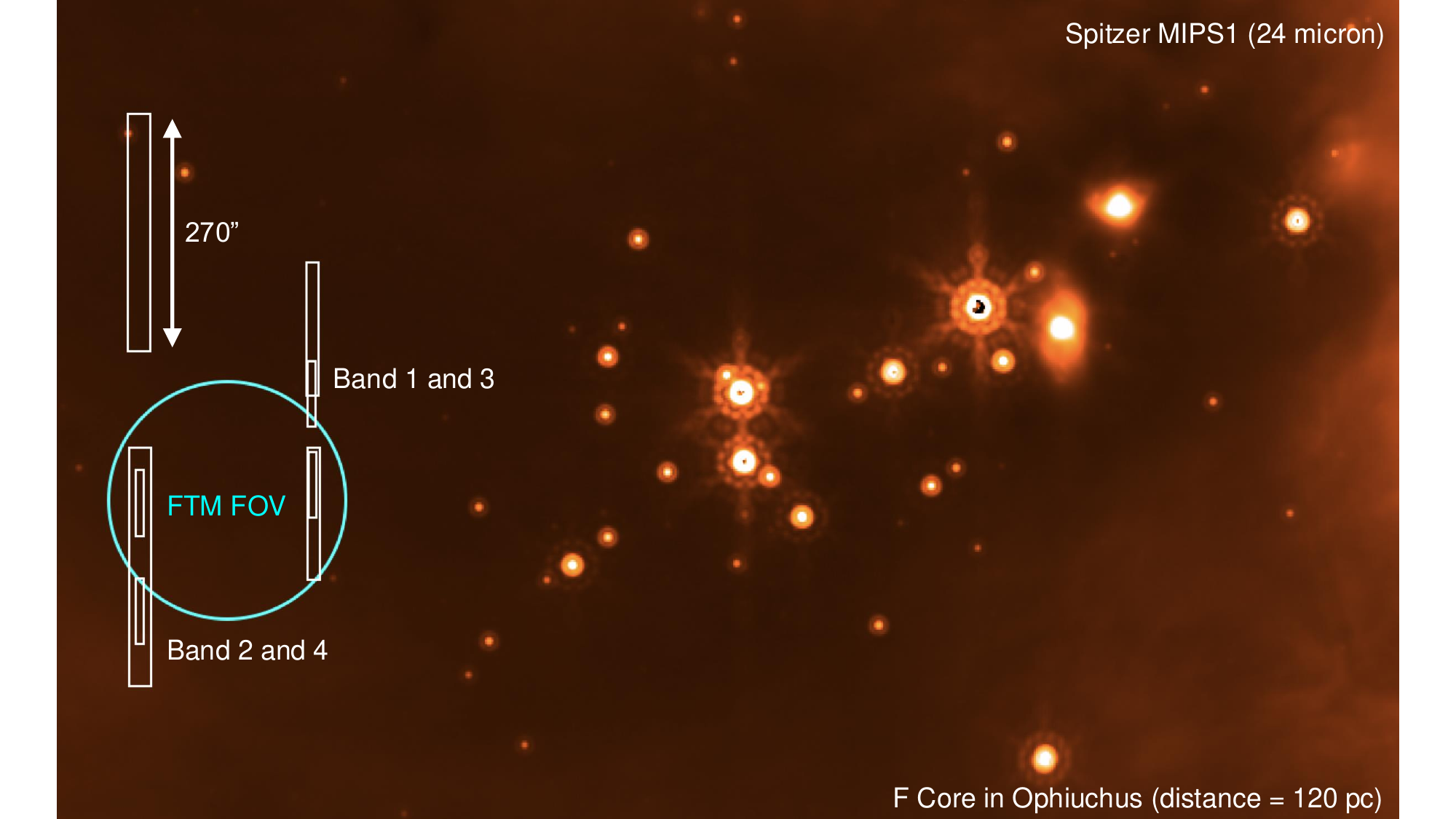}
\caption{The spatial sampling and resolution of FIRESS is well-suited to observe individual young star and protoplanetary disks in nearby star-forming regions and young clusters. The FIRESS field of view compared to a 24\,$\mu$m Spitzer MIPS image of the protostellar cluster in Ophiuchus F core at a distance of 120\,pc. This is typical a nearby young star-forming region likely to be targeted by PRIMA with a spatial resolution corresponding to that of PRIMA at $\sim 50\,\mu$m. The image is centered near RA=16h27m28s, Dec=-24d40m10s.}
\label{fig:fov}
\end{figure}

The potential for source confusion with a single-dish far-infrared telescope is an important consideration. However, the layout of the FIRESS slits and field of view is well matched to observe individual protoplanetary disks in nearby clusters of young stars. With spatial resolutions of 7.6-22.9'', depending on spectral channel, FIRESS can obtain spectra of young stars with projected separations characteristic of the clustering regime \citep[$\gtrsim$5000\,AU or $\gtrsim$ 30'' at the distance of the Taurus star-forming region][]{Simon97,Kraus08}. Figure \ref{fig:fov} shows the FIRESS FOV compared to one of the densest regions in the Ophiuchus star-forming regions. This is a young cluster, and since stellar clusters disperse with age this represents a stressing case for PRIMA, as other clusters will have lower angular density. The low-resolution mapping mode with PRIMA will be able to effectively separate protoplanetary disks in young clusters out to the distance of Orion (460\,pc).  

\subsection{Resolving power}

The resolving power is set to be sufficient to detect relevant lines with adequate line-to-continuum ratios over the bright dust continuum in galaxies using the low-resolution grating, and from typical protoplanetary disks using the FTM, without sacrificing the broad wavelength coverage or sensitivity \citep[see][for the FIRESS technology trade study]{Bradford25}. For the low-resolution mode, a  resolving power of $R>80$ at all wavelengths ensures line-to-continuum ratios of the [Ne II] and [O IV] lines in excess of 3\% \citep{Bonato19, Gruppioni16} for luminous galaxies with AGN luminosity fraction $F_{\rm AGN}>0.4$. 

\begin{figure}[ht]
\centering
\includegraphics[width=10cm]{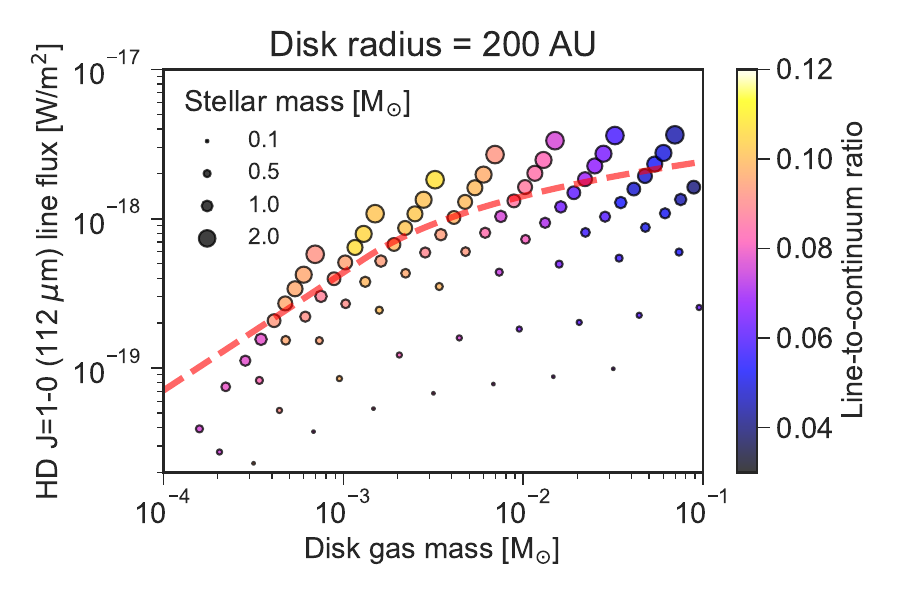}
\caption{Predicted HD 1-0 line fluxes and line-to-continuum ratios for a representative grid of disk models using the model described in \citep{Blevins16}. The line flux predictions and their detailed correlation with disk gas masses have been independently confirmed by at least three different studies \citep{McClure16,Trapman17,Seo24}; the red dashed curve is from \citep{Trapman17}.}
\label{fig:resolving}
\end{figure}

The resolving power of the high-resolution mode is designed to allow the 112\,$\mu$m HD 1-0 line to have a contrast (line-to-continuum ratio) of at least 2.5\% for all disks more massive than $1\,M_{\rm Jup}$, as predicted by available radiative transfer models of protoplanetary disks \citep{McClure16,Trapman17,Seo24}. Given variation in model predictions, this translates to resolving powers of at least $R\sim 3000$ to ensure sufficient line-to-continuum ratios at 112\,$\mu$m for a wide range of disk masses and gas-to-dust ratios (see Figure \ref{fig:resolving}). It is further required that models are consistent with the existing detections of the HD line in three disks (TW Hya, DM Tau, and GM Aur) obtained by Herschel-PACS \citep{Bergin13,McClure16}. These are observed to have line-to-continuum ratios at $R\sim 3000$ of 12\%, 6\%, and 40\%, respectively. 

Since the resolving power of the FTM scales as $R\sim 4400 \times (112\,\rm \mu m)/\lambda$, the highest power anticipated is achieved at the shortest wavelengths, with $R(24\,\rm \mu m) =20,500$ ($15\,\rm km\,s^{-1}$). This is sufficient to kinematically resolve water lines at the snowline typically across several resolution elements (the full width of a line at the surface snowline around a solar-mass star is $\sim 35\,\rm km\,s^{-1}$). At the longer end of the wavelength range, at 235\,$\mu$m, the resolving power is at least $R\sim 2000$, sufficient to maintain line-to-continuum ratios of the water ground-state line in excess of a few \% \citep{Blevins16}. 

As the high-resolution mode of FIRESS is implemented by a Fourier Transform Module (FTM), the resolving power may be tuned to the specific science case by changing the maximum path difference in a spectral scan, trading against sensitivity. This is used to optimize FIRESS observations of galactic outflows (Figure \ref{fig:oh}).

\subsection{Spectroscopic sensitivity and survey speed}
\label{section:speed}
\begin{figure}[ht]
\centering
\includegraphics[width=10cm]{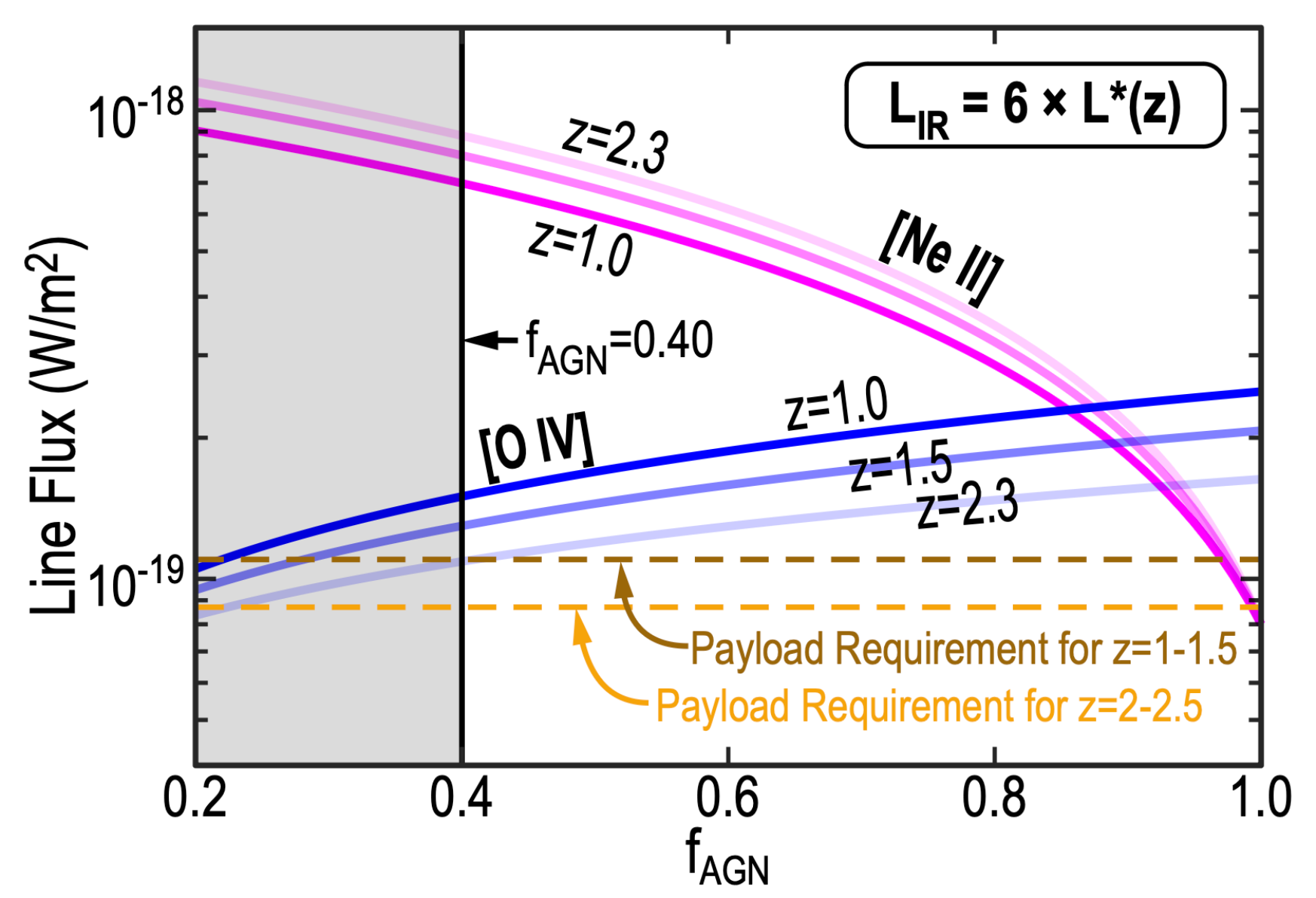}
\caption{Relative fine structure line fluxes tracing star-formation ([Ne II]) and AGN activity ([O IV]) from a luminous star-forming galaxy as a function of AGN fraction, compared to the FIRESS sensitivity. FIRESS can detect an AGN fraction ($f_{\rm AGN}$ greater than 0.4 at cosmic noon.}
\label{fig:smbh_sensitivity}
\end{figure}

The spectroscopic sensitivity of the FIRESS low-resolution mode is sufficient to detect PAH emission from luminous ($\gtrsim 10^{11}\,L_{\odot}$) galaxies over a wide range of redshifts, out to cosmic noon (z$\lesssim$2, Figure \ref{fig:galaxy spectra}). For pointed spectroscopy, this enables the detection of line emission from large numbers of luminous galaxies in the distant universe (see Figure \ref{fig:smbh_sensitivity}). The FIRESS low-resolution mode can be operated in a fast-scan pattern to obtain $R\sim 80-130$ spectral imaging of large areas on the sky (e.g., 1 sq. degree to 5$\sigma$ line detections better than $5\times 10^{-18}\,\rm W m^{-2}$ in 100 hours). 

The broad-band coverage of the high-resolution FTM offers dramatic improvements in the speed with which the spectral dimension is covered. For instance, for full spectral range scans (51-210\,$\mu$m) Herschel-PACS reached a line RMS of 0.45-3\,$\times 10^{-17}\,\rm W/m^2$ in 1 hour \citep{Green13}. The same coverage can be covered by FIRESS to an RMS of at least 1.4\,$\times 10^{-19}\,\rm W/m^2$, or 30-200 times the sensitivity of PACS in the same time, depending on wavelength. This corresponds to an improvement of 3-4 orders of magnitude in survey speed over more than twice the wavelength coverage, at 3-5 times the spectral resolving power, compared to the previous state-of-the-art. 

For the high-resolution mode, an important consideration is that the FTM will receive a noise penalty from the integrated background across a full low-resolution pixel. This results in a moderate sensitivity penalty for the brightest protoplanetary disks with dust continua brighter than $\sim$0.3\,Jy range, while fainter disks are much less affected. Essentially, the detectable line-to-continuum ratio is only a weak function of source brightness in the bright-source limit. The trade-off is that the entire spectrum is measured at high resolution, offering an expansive discovery space for a wide range of scientific applications that would not otherwise be possible. 

\subsection{Discovery space and general observer science}
\label{sec:discover}
The highly multiplexed FIRESS modes have potential for a wide range of science cases and can address objectives that may emerge in the next decade particularly leveraging  JWST and ALMA. In the PRIMA GO science book, 2/3 of the science cases specify the use of FIRESS, with roughly equal representation of the different FIRESS observing modes. Highlights from the PRIMA GO science book \citep{Moullet23} include sensitive measurements of the D/H ratio in solar system comets to elucidate the origin of Earth's water (Lis et al.), determining the primary modes of protostellar accretion by monitoring large numbers of protostars (Battersby et al.), detecting the primary reservoirs of nitrogen in protoplanetary disks (Bergner et al.), the evolution of crystalline silicates at cosmic noon (Kemper et al.), searching for the missing oxygen in the interstellar medium using far-infrared ice bands and [OI] emission (Onaka et al.), understanding the energy balance in nearby galaxies using fine-structure lines (Sutter et al.), and many more. We anticipate gathering more science cases for PRIMA FIRESS during phase A and beyond. 

\subsection*{Disclosures}
The authors have no financial interests in the manuscript or other potential conflicts of interest to disclose. 


\subsection* {Code, Data, and Materials Availability} 
The PRIMA exposure time calculator is available on {\tt prima.ipac.caltech.edu}. This paper uses data available in ESA's Infrared Space Observatory and Herschel Science Data Archives and the Spitzer Archive at IRSA.


\subsection* {Acknowledgments}
This research was carried out at the Jet Propulsion Laboratory, California Institute of Technology, under a contract with the National Aeronautics and Space Administration (80NM0018D0004).


\bibliography{report}   

\begin{thebibliography}{10}

\bibitem{Glenn25}
J.~{Glenn}, ``{PRIMA mission concept},'' {\em JATIS} {\bf 11}(3)  (2025).

\bibitem{Fixsen98}
D.~J. {Fixsen}, E.~{Dwek}, J.~C. {Mather}, {\em et~al.}, ``{The Spectrum of the Extragalactic Far-Infrared Background from the COBE FIRAS Observations},'' {\em \apj} {\bf 508}, 123--128  (1998).

\bibitem{Hill18}
R.~{Hill}, K.~W. {Masui}, and D.~{Scott}, ``{The Spectrum of the Universe},'' {\em Applied Spectroscopy} {\bf 72}, 663--688  (2018).

\bibitem{Ciesla25}
L.~{Ciesla}, ``{PRIMA: PRIMAger, a far-infrared hyperspectral and polarimetric instrument},'' {\em JATIS} {\bf 11}(031625)  (2025).

\bibitem{Bradford25}
M.~{Bradford}, ``{},'' {\em JATIS} {\bf TBD}(TBD)  (2025).

\bibitem{Kessler96}
M.~F. {Kessler}, J.~A. {Steinz}, M.~E. {Anderegg}, {\em et~al.}, ``{The Infrared Space Observatory (ISO) mission.},'' {\em \aap} {\bf 315}, L27--L31  (1996).

\bibitem{Pilbratt10}
G.~L. {Pilbratt}, J.~R. {Riedinger}, T.~{Passvogel}, {\em et~al.}, ``{Herschel Space Observatory. An ESA facility for far-infrared and submillimetre astronomy},'' {\em \aap} {\bf 518}, L1  (2010).

\bibitem{Becklin07}
E.~E. {Becklin}, A.~G.~G.~M. {Tielens}, R.~D. {Gehrz}, {\em et~al.}, ``{Stratospheric Observatory for Infrared Astronomy (SOFIA)},'' in {\em Infrared Spaceborne Remote Sensing and Instrumentation XV},  M.~{Strojnik-Scholl}, Ed., {\em Society of Photo-Optical Instrumentation Engineers (SPIE) Conference Series} {\bf 6678}, 66780A  (2007).

\bibitem{Werner04}
M.~W. {Werner}, T.~L. {Roellig}, F.~J. {Low}, {\em et~al.}, ``{The Spitzer Space Telescope Mission},'' {\em \apjs} {\bf 154}, 1--9  (2004).

\bibitem{Gardner06}
J.~P. {Gardner}, J.~C. {Mather}, M.~{Clampin}, {\em et~al.}, ``{The James Webb Space Telescope},'' {\em \ssr} {\bf 123}, 485--606  (2006).

\bibitem{Clegg96}
P.~E. {Clegg}, P.~A.~R. {Ade}, C.~{Armand}, {\em et~al.}, ``{The ISO Long-Wavelength Spectrometer.},'' {\em \aap} {\bf 315}, L38--L42  (1996).

\bibitem{Swinyard96}
B.~M. {Swinyard}, P.~E. {Clegg}, P.~A.~R. {Ade}, {\em et~al.}, ``{Calibration and performance of the ISO Long-Wavelength Spectrometer.},'' {\em \aap} {\bf 315}, L43--L48  (1996).

\bibitem{Poglitsch10}
A.~{Poglitsch}, C.~{Waelkens}, N.~{Geis}, {\em et~al.}, ``{The Photodetector Array Camera and Spectrometer (PACS) on the Herschel Space Observatory},'' {\em \aap} {\bf 518}, L2  (2010).

\bibitem{Heyminck12}
S.~{Heyminck}, U.~U. {Graf}, R.~{G{\"u}sten}, {\em et~al.}, ``{GREAT: the SOFIA high-frequency heterodyne instrument},'' {\em \aap} {\bf 542}, L1  (2012).

\bibitem{Moullet23}
A.~{Moullet}, T.~{Kataria}, D.~{Lis}, {\em et~al.}, ``{PRIMA General Observer Science Book},'' {\em arXiv e-prints} , arXiv:2310.20572  (2023).

\bibitem{Mordasini16}
C.~{Mordasini}, R.~{van Boekel}, P.~{Molli{\`e}re}, {\em et~al.}, ``{The Imprint of Exoplanet Formation History on Observable Present-day Spectra of Hot Jupiters},'' {\em \apj} {\bf 832}, 41  (2016).

\bibitem{Oberg11}
K.~I. {{\"O}berg}, R.~{Murray-Clay}, and E.~A. {Bergin}, ``{The Effects of Snowlines on C/O in Planetary Atmospheres},'' {\em \apjl} {\bf 743}, L16  (2011).

\bibitem{Madhusudhan12}
N.~{Madhusudhan}, ``{C/O Ratio as a Dimension for Characterizing Exoplanetary Atmospheres},'' {\em \apj} {\bf 758}, 36  (2012).

\bibitem{Eistrup18}
C.~{Eistrup}, C.~{Walsh}, and E.~F. {van Dishoeck}, ``{Molecular abundances and C/O ratios in chemically evolving planet-forming disk midplanes},'' {\em \aap} {\bf 613}, A14  (2018).

\bibitem{Bean18}
J.~L. {Bean}, K.~B. {Stevenson}, N.~M. {Batalha}, {\em et~al.}, ``{The Transiting Exoplanet Community Early Release Science Program for JWST},'' {\em \pasp} {\bf 130}, 114402  (2018).

\bibitem{Constantinou23}
S.~{Constantinou}, N.~{Madhusudhan}, and S.~{Gandhi}, ``{Early Insights for Atmospheric Retrievals of Exoplanets Using JWST Transit Spectroscopy},'' {\em \apjl} {\bf 943}, L10  (2023).

\bibitem{Ros13}
K.~{Ros} and A.~{Johansen}, ``{Ice condensation as a planet formation mechanism},'' {\em \aap} {\bf 552}, A137  (2013).

\bibitem{Lambrechts19}
M.~{Lambrechts}, A.~{Morbidelli}, S.~A. {Jacobson}, {\em et~al.}, ``{Formation of planetary systems by pebble accretion and migration. How the radial pebble flux determines a terrestrial-planet or super-Earth growth mode},'' {\em \aap} {\bf 627}, A83  (2019).

\bibitem{Pinilla20}
P.~{Pinilla}, I.~{Pascucci}, and S.~{Marino}, ``{Hints on the origins of particle traps in protoplanetary disks given by the M$_{dust}$ - M$_{{\ensuremath{\star}}}$ relation},'' {\em \aap} {\bf 635}, A105  (2020).

\bibitem{Pinilla12}
P.~{Pinilla}, M.~{Benisty}, and T.~{Birnstiel}, ``{Ring shaped dust accumulation in transition disks},'' {\em \aap} {\bf 545}, A81  (2012).

\bibitem{Appelgren25}
J.~{Appelgren}, A.~{Johansen}, M.~{Lambrechts}, {\em et~al.}, ``{The evolution of the flux-size relationship in protoplanetary discs by viscous evolution and radial pebble drift},'' {\em arXiv e-prints} , arXiv:2501.04411  (2025).

\bibitem{Banzatti23}
A.~{Banzatti}, K.~M. {Pontoppidan}, J.~S. {Carr}, {\em et~al.}, ``{JWST Reveals Excess Cool Water near the Snow Line in Compact Disks, Consistent with Pebble Drift},'' {\em \apjl} {\bf 957}, L22  (2023).

\bibitem{Romero-Mirza24}
C.~E. {Romero-Mirza}, A.~{Banzatti}, K.~I. {{\"O}berg}, {\em et~al.}, ``{Retrieval of Thermally Resolved Water Vapor Distributions in Disks Observed with JWST-MIRI},'' {\em \apj} {\bf 975}, 78  (2024).

\bibitem{Colmenares24}
M.~J. {Colmenares}, E.~A. {Bergin}, C.~{Salyk}, {\em et~al.}, ``{JWST/MIRI Detection of a Carbon-rich Chemistry in the Disk of a Solar Nebula Analog},'' {\em \apj} {\bf 977}, 173  (2024).

\bibitem{Kanwar24}
J.~{Kanwar}, I.~{Kamp}, H.~{Jang}, {\em et~al.}, ``{MINDS. Hydrocarbons detected by JWST/MIRI in the inner disk of Sz28 consistent with a high C/O gas-phase chemistry},'' {\em \aap} {\bf 689}, A231  (2024).

\bibitem{Long25}
F.~{Long}, I.~{Pascucci}, A.~{Houge}, {\em et~al.}, ``{The First JWST View of a 30-Myr-old Protoplanetary Disk Reveals a Late-stage Carbon-rich Phase},'' {\em \apjl} {\bf 978}, L30  (2025).

\bibitem{Houge25}
A.~{Houge}, S.~{Krijt}, A.~{Banzatti}, {\em et~al.}, ``{Smuggling unnoticed: Towards a 2D view of water and dust delivery to the inner regions of protoplanetary discs},'' {\em \mnras}   (2025).

\bibitem{Bruderer12}
S.~{Bruderer}, E.~F. {van Dishoeck}, S.~D. {Doty}, {\em et~al.}, ``{The warm gas atmosphere of the HD 100546 disk seen by Herschel. Evidence of a gas-rich, carbon-poor atmosphere?},'' {\em \aap} {\bf 541}, A91  (2012).

\bibitem{Bergin19}
E.~{Bergin}, K.~{Pontoppidan}, C.~{Bradford}, {\em et~al.}, ``{The Disk Gas Mass and the Far-IR Revolution},'' {\em \baas} {\bf 51}, 222  (2019).

\bibitem{Linsky98}
J.~L. {Linsky}, ``{Deuterium Abundance in the Local ISM and Possible Spatial Variations},'' {\em \ssr} {\bf 84}, 285--296  (1998).

\bibitem{Reines2015}
A.~E. {Reines} and M.~{Volonteri}, ``{Relations between Central Black Hole Mass and Total Galaxy Stellar Mass in the Local Universe},'' {\em ApJ} {\bf 813}, 82  (2015).

\bibitem{Zavala2021}
J.~A. {Zavala}, C.~M. {Casey}, S.~M. {Manning}, {\em et~al.}, ``{The Evolution of the IR Luminosity Function and Dust-obscured Star Formation over the Past 13 Billion Years},'' {\em ApJ} {\bf 909}, 165  (2021).

\bibitem{Hickox2018}
R.~C. {Hickox} and D.~M. {Alexander}, ``{Obscured Active Galactic Nuclei},'' {\em ARA\&A} {\bf 56}, 625--671  (2018).

\bibitem{Koss2011}
M.~{Koss}, R.~{Mushotzky}, S.~{Veilleux}, {\em et~al.}, ``{Host Galaxy Properties of the Swift Bat Ultra Hard X-Ray Selected Active Galactic Nucleus},'' {\em ApJ} {\bf 739}, 57  (2011).

\bibitem{Comastri2004}
A.~{Comastri}, ``{Compton-Thick AGN: The Dark Side of the X-Ray Background},'' in {\em Supermassive Black Holes in the Distant Universe},  A.~J. {Barger}, Ed., {\em Astrophysics and Space Science Library} {\bf 308}, 245  (2004).

\bibitem{Begelman1979}
M.~C. {Begelman}, ``{Can a spherically accreting black hole radiate very near the Eddington limit?},'' {\em MNRAS} {\bf 187}, 237--251  (1979).

\bibitem{Pacucci2024}
F.~{Pacucci} and R.~{Narayan}, ``{Mildly Super-Eddington Accretion onto Slowly Spinning Black Holes Explains the X-Ray Weakness of the Little Red Dots},'' {\em ApJ} {\bf 976}, 96  (2024).

\bibitem{Kirkpatrick2013}
A.~{Kirkpatrick}, A.~{Pope}, V.~{Charmandaris}, {\em et~al.}, ``{GOODS-Herschel: Separating High-redshift Active Galactic Nuclei and Star-forming Galaxies Using Infrared Color Diagnostics},'' {\em ApJ} {\bf 763}, 123  (2013).

\bibitem{Shimizu2017}
T.~T. {Shimizu}, R.~F. {Mushotzky}, M.~{Mel{\'e}ndez}, {\em et~al.}, ``{Herschel far-infrared photometry of the Swift Burst Alert Telescope active galactic nuclei sample of the local universe - III. Global star-forming properties and the lack of a connection to nuclear activity},'' {\em MNRAS} {\bf 466}, 3161--3183  (2017).

\bibitem{Mitchell22}
P.~D. {Mitchell} and J.~{Schaye}, ``{How gas flows shape the stellar-halo mass relation in the EAGLE simulation},'' {\em \mnras} {\bf 511}, 2948--2967  (2022).

\bibitem{Nelson19}
D.~{Nelson}, A.~{Pillepich}, V.~{Springel}, {\em et~al.}, ``{First results from the TNG50 simulation: galactic outflows driven by supernovae and black hole feedback},'' {\em \mnras} {\bf 490}, 3234--3261  (2019).

\bibitem{Veilleux20}
S.~{Veilleux}, R.~{Maiolino}, A.~D. {Bolatto}, {\em et~al.}, ``{Cool outflows in galaxies and their implications},'' {\em \aapr} {\bf 28}, 2  (2020).

\bibitem{Matsuura09}
M.~{Matsuura}, M.~J. {Barlow}, A.~A. {Zijlstra}, {\em et~al.}, ``{The global gas and dust budget of the Large Magellanic Cloud: AGB stars and supernovae, and the impact on the ISM evolution},'' {\em \mnras} {\bf 396}, 918--934  (2009).

\bibitem{Nittler03}
L.~R. {Nittler}, ``{Presolar stardust in meteorites: recent advances and scientific frontiers},'' {\em Earth and Planetary Science Letters} {\bf 209}, 259--273  (2003).

\bibitem{Boyer12}
M.~L. {Boyer}, S.~{Srinivasan}, D.~{Riebel}, {\em et~al.}, ``{The Dust Budget of the Small Magellanic Cloud: Are Asymptotic Giant Branch Stars the Primary Dust Source at Low Metallicity?},'' {\em \apj} {\bf 748}, 40  (2012).

\bibitem{Popping17}
G.~{Popping}, R.~S. {Somerville}, and M.~{Galametz}, ``{The dust content of galaxies from z = 0 to z = 9},'' {\em \mnras} {\bf 471}, 3152--3185  (2017).

\bibitem{Peroux20}
C.~{P{\'e}roux} and J.~C. {Howk}, ``{The Cosmic Baryon and Metal Cycles},'' {\em \araa} {\bf 58}, 363--406  (2020).

\bibitem{Smith07}
J.~D.~T. {Smith}, B.~T. {Draine}, D.~A. {Dale}, {\em et~al.}, ``{The Mid-Infrared Spectrum of Star-forming Galaxies: Global Properties of Polycyclic Aromatic Hydrocarbon Emission},'' {\em \apj} {\bf 656}, 770--791  (2007).

\bibitem{Langeroodi23}
D.~{Langeroodi} and J.~{Hjorth}, ``{Little Red Dots or Brown Dwarfs? NIRSpec Discovery of Three Distant Brown Dwarfs Masquerading as NIRCam-selected Highly Reddened Active Galactic Nuclei},'' {\em \apjl} {\bf 957}, L27  (2023).

\bibitem{Markov23}
V.~{Markov}, S.~{Gallerani}, A.~{Pallottini}, {\em et~al.}, ``{Dust attenuation law in JWST galaxies at z {\ensuremath{\sim}} 7-8},'' {\em \aap} {\bf 679}, A12  (2023).

\bibitem{Spilker23}
J.~S. {Spilker}, K.~A. {Phadke}, M.~{Aravena}, {\em et~al.}, ``{Spatial variations in aromatic hydrocarbon emission in a dust-rich galaxy},'' {\em \nat} {\bf 618}, 708--711  (2023).

\bibitem{Engelbracht05}
C.~W. {Engelbracht}, K.~D. {Gordon}, G.~H. {Rieke}, {\em et~al.}, ``{Metallicity Effects on Mid-Infrared Colors and the 8 {\ensuremath{\mu}}m PAH Emission in Galaxies},'' {\em \apjl} {\bf 628}, L29--L32  (2005).

\bibitem{Engelbracht08}
C.~W. {Engelbracht}, G.~H. {Rieke}, K.~D. {Gordon}, {\em et~al.}, ``{Metallicity Effects on Dust Properties in Starbursting Galaxies},'' {\em \apj} {\bf 678}, 804--827  (2008).

\bibitem{Whitcomb23}
C.~M. {Whitcomb}, K.~{Sandstrom}, A.~{Leroy}, {\em et~al.}, ``{Star Formation and Molecular Gas Diagnostics with Mid- and Far-infrared Emission},'' {\em \apj} {\bf 948}, 88  (2023).

\bibitem{Colbert99}
J.~W. {Colbert}, M.~A. {Malkan}, P.~E. {Clegg}, {\em et~al.}, ``{ISO LWS Spectroscopy of M82: A Unified Evolutionary Model},'' {\em \apj} {\bf 511}, 721--729  (1999).

\bibitem{Sturm00}
E.~{Sturm}, D.~{Lutz}, D.~{Tran}, {\em et~al.}, ``{ISO-SWS spectra of galaxies: Continuum and features},'' {\em \aap} {\bf 358}, 481--493  (2000).

\bibitem{Kaufman06}
M.~J. {Kaufman}, M.~G. {Wolfire}, and D.~J. {Hollenbach}, ``{[Si II], [Fe II], [C II], and H$_{2}$ Emission from Massive Star-forming Regions},'' {\em \apj} {\bf 644}, 283--299  (2006).

\bibitem{Pontoppidan24}
K.~M. {Pontoppidan}, C.~{Salyk}, A.~{Banzatti}, {\em et~al.}, ``{High-contrast JWST-MIRI Spectroscopy of Planet-forming Disks for the JDISC Survey},'' {\em \apj} {\bf 963}, 158  (2024).

\bibitem{Blevins16}
S.~M. {Blevins}, K.~M. {Pontoppidan}, A.~{Banzatti}, {\em et~al.}, ``{Measurements of Water Surface Snow Lines in Classical Protoplanetary Disks},'' {\em \apj} {\bf 818}, 22  (2016).

\bibitem{Malfait99}
K.~{Malfait}, C.~{Waelkens}, J.~{Bouwman}, {\em et~al.}, ``{The ISO spectrum of the young star HD 142527},'' {\em \aap} {\bf 345}, 181--186  (1999).

\bibitem{Sturm13}
B.~{Sturm}, J.~{Bouwman}, T.~{Henning}, {\em et~al.}, ``{The 69 {\ensuremath{\mu}}m forsterite band in spectra of protoplanetary disks. Results from the Herschel DIGIT programme},'' {\em \aap} {\bf 553}, A5  (2013).

\bibitem{Kamp18}
I.~{Kamp}, A.~{Scheepstra}, M.~{Min}, {\em et~al.}, ``{Diagnostic value of far-IR water ice features in T Tauri disks},'' {\em \aap} {\bf 617}, A1  (2018).

\bibitem{Simon97}
M.~{Simon}, ``{Clustering of Young Stars in Taurus, Ophiuchus, and the Orion Trapezium},'' {\em \apjl} {\bf 482}, L81--L84  (1997).

\bibitem{Kraus08}
A.~L. {Kraus} and L.~A. {Hillenbrand}, ``{Spatial Distributions of Young Stars},'' {\em \apjl} {\bf 686}, L111  (2008).

\bibitem{Bonato19}
M.~{Bonato}, G.~{De Zotti}, D.~{Leisawitz}, {\em et~al.}, ``{Origins Space Telescope: Predictions for far-IR spectroscopic surveys},'' {\em \pasa} {\bf 36}, e017  (2019).

\bibitem{Gruppioni16}
C.~{Gruppioni}, S.~{Berta}, L.~{Spinoglio}, {\em et~al.}, ``{Tracing black hole accretion with SED decomposition and IR lines: from local galaxies to the high-z Universe},'' {\em \mnras} {\bf 458}, 4297--4320  (2016).

\bibitem{McClure16}
M.~K. {McClure}, E.~A. {Bergin}, L.~I. {Cleeves}, {\em et~al.}, ``{Mass Measurements in Protoplanetary Disks from Hydrogen Deuteride},'' {\em \apj} {\bf 831}, 167  (2016).

\bibitem{Trapman17}
L.~{Trapman}, A.~{Miotello}, M.~{Kama}, {\em et~al.}, ``{Far-infrared HD emission as a measure of protoplanetary disk mass},'' {\em \aap} {\bf 605}, A69  (2017).

\bibitem{Seo24}
Y.~M. {Seo}, K.~{Willacy}, G.~{Bryden}, {\em et~al.}, ``{Retrievals of Protoplanetary Disk Parameters Using Thermochemical Models. I. Disk Gas Mass from Hydrogen Deuteride Spectroscopy},'' {\em \apj} {\bf 967}, 131  (2024).

\bibitem{Bergin13}
E.~A. {Bergin}, L.~I. {Cleeves}, U.~{Gorti}, {\em et~al.}, ``{An old disk still capable of forming a planetary system},'' {\em \nat} {\bf 493}, 644--646  (2013).

\bibitem{Green13}
J.~D. {Green}, N.~J. {Evans}, II, J.~K. {J{\o}rgensen}, {\em et~al.}, ``{Embedded Protostars in the Dust, Ice, and Gas In Time (DIGIT) Herschel Key Program: Continuum SEDs, and an Inventory of Characteristic Far-infrared Lines from PACS Spectroscopy},'' {\em \apj} {\bf 770}, 123  (2013).

\end{thebibliography}
\bibliographystyle{spiejour}   





\end{document}